\documentclass[twocolumn,twoside,slac_two]{revtex4}
\usepackage{graphicx}
\usepackage{fancyhdr}
\begin{document}

\title{Swift/BAT monitoring of Fermi/LAT sources}

%

%

\author{ Hans A. Krimm}
\affiliation{CRESST, Universities Space Research Association  and NASA GSFC, Greenbelt,  MD 20771, USA}

\author{Scott D. Barthelmy, Neil Gehrels,  Jack Tueller}
\affiliation{NASA Goddard Space Flight Center, Greenbelt,
  MD 20771, USA}

\author{Wayne H. Baumgartner, Jay R. Cummings,  Taka Sakamoto}
\affiliation{CRESST, University of Maryland, Baltimore County  and NASA GSFC, Greenbelt,  MD 20771, USA}

\author{Edward E. Fenimore, David M. Palmer}
\affiliation{Los Alamos National Laboratory, P.O. Box 1663, Los Alamos, NM 87545, USA}

\author{Craig B. Markwardt, Gerald K. Skinner}
\affiliation{CRESST, University of Maryland, College Park and NASA GSFC, Greenbelt,  MD 20771, USA}

\author{Michael  Stamatikos}
\affiliation{NASA Goddard Space Flight Center and Ohio State University, Columbus, OH 43210, USA}


\begin{abstract}
The Swift Burst Alert Telescope (BAT) hard X-ray transient monitor tracks more than 700 galactic and extragalactic
sources on time scales ranging from a single Swift pointing (approximately 20 minutes) to one day.
The monitored sources include all objects from the Fermi LAT bright source list which are either identified
or which have a 95\% error confidence radius of less than eight arc minutes. We report on the detection
statistics of these sources in the BAT monitor both before and after the launch of Fermi.

\end{abstract}

\maketitle

\thispagestyle{fancy}


\section{The {\em Swift}/BAT Hard X-ray Survey}
\noindent The {\em Swift}/BAT Hard X-ray Survey~\cite{tuel09,survey-web} uses the full {\em Swift}/BAT data in eight energy ranges (covering 14-195~keV) to produce a map (mosaic) of the full sky.  The primary goal of the BAT Survey is to make a complete census of nearby AGN down to a $4.8\sigma$\ sensitivity  of $2.2 \times 10^{-11} erg\ cm^{-2} s^{-1}$ (1 mCrab).  The current published results cover the first 22~months of the {\em Swift} mission from December 2004 through September 2006.  The data used in this paper cover the first 36~months (through November 2007) and processing is underway to complete the full 60~months of the mission to date.  Although sources are only announced if they are at least $4.8\sigma$, we have used the sky maps to search for lower level emission from the 175 sources in the LAT Bright Source list which are included in the BAT Monitor catalog.
	Other work which correlates BAT and LAT blazar detections has been done by Sambruna et al (in preparation) who have used BAT and LAT spectral fits to argue in favor of the ``blazar sequence" relating the X-ray and $\gamma$-ray continua.

\section{Summary of Results}

\begin{itemize}

\item {Using a low threshold, we find that BAT can detect 31 of the 108 identified blazars in the LAT Bright Source Catalog and 18 other sources}
\item {Given the significance distribution, we expect that many more blazars will be detected as the BAT Survey and Monitor become more sensitive}
\item {Preliminary blazar SEDs in the {\em Swift} energy range are consistent with the standard models of FSRQ and BL Lac blazar types}

\end{itemize}

\section{Distribution of BAT Significances}

\begin{figure*}[t]
\centering
\includegraphics[width=135mm]{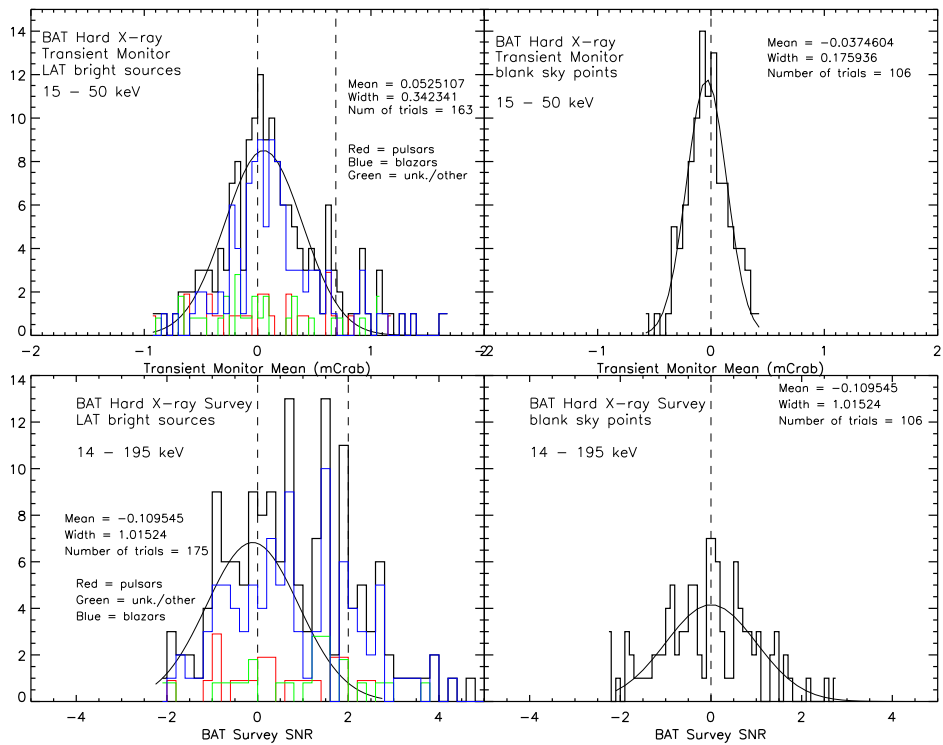} 
 \caption{Histograms of count rates and signal-to-noise ratios (SNR) for {\em Fermi}/LAT bright sources from the {\em Swift}/BAT Hard X-ray Transient Monitor and Hard X-ray Survey.   {\bf Top Left:} Histogram of the average count rate derived for each {\em Fermi}/LAT bright source in the BAT Transient Monitor. 
Note the asymmetry in the distribution.  The dashed vertical line to the right indicates the detection threshold of $2\sigma$.  {\bf Top Right:} A similar distribution for ``blank sky" points in the monitor catalog.  The distribution is narrower than the one on the left because these sources were monitored for a longer period.  {\bf Bottom Left:}  A distribution of SNR from the BAT Survey, again showing a pronounced positive excess.  {\bf Bottom Right:} The distribution for the blank sky points shows no such asymmetry. 
}\label{fig1}
\end{figure*}

\noindent The histograms in Figure~\ref{fig1} are derived from the  {\em Swift}/BAT Hard X-ray Transient Monitor (top) and  {\em Swift}/BAT Hard X-ray Survey (bottom).   The plots on the left show sources in the first {\em Fermi}/LAT Bright Source List and the plots on the right show sources from 106 ``blank sky" points in the Transient Monitor catalog.  The blank sky points are distributed randomly across the sky and each point has been checked against all of the X-ray catalogs in Simbad (http://simbad.u-strasbg.fr/simbad/sim-fid) to ensure that there is no hard or soft X-ray source nearby.  

	To create the Transient Monitor histograms, we have calculated the mean daily average count rate for each source and divided by the mean count rate for a source one thousandth the rate of the Crab ($2.22 \times 10^{-4}\ ct\ cm^{-2} s^{-1}$).  The histograms are color-coded for source type, with black representing the sum of the three categories.  For non-detections, these means are centered on zero (top right of Fig~\ref{fig1}).  By contrast, for the LAT bright sources, there is a strong bias toward the positive side, telling us that a high percentage of these sources are detected.  In the top left of Fig~\ref{fig1} there are 35 sources at $> 2\sigma$ (including 12 off the plot to the right), compared to 4 expected.  This enables us to set a low detection threshold, as indicated by the dashed vertical line to the right, corresponding to 0.68 mCrab.  The width of the blank sky distribution is smaller than that of the {\em Fermi} bright sources for the following reason.  Light curves for the blank sky points have been accumulated since near the start of the {\em Swift} mission (57 months since Feb. 2005), while light curves for most of the LAT sources have only been accumulated for only 9 months since the release of the first LAT bright source catalog.  Thus the distribution of means is tighter.  

	A similar pair of histograms are created from the results of the 36-month {\em Swift}/BAT Hard X-ray Survey\cite{tuel10}.  In these plots, the histograms are based on the signal-to-noise ratio (SNR) for each source.  Therefore, the blank-sky histogram has a mean near zero and width near unity.  The fit to the blank-sky histogram is overlaid on the bright-source histogram.  For the survey, with a sensitivity about twice that of the monitor (due mostly to the longer collection time), we see an even greater asymmetry to the positive side.  This suggests that a large number of the LAT blazars are in fact detected in the BAT and that many more will be detected as the Survey and Monitor run longer and become more sensitive.  The dashed vertical line  to the right indicates $2\sigma$.



	\section{Relation of BAT and LAT Fluxes}

\begin{figure*}[t]
\centering
\includegraphics[width=135mm]{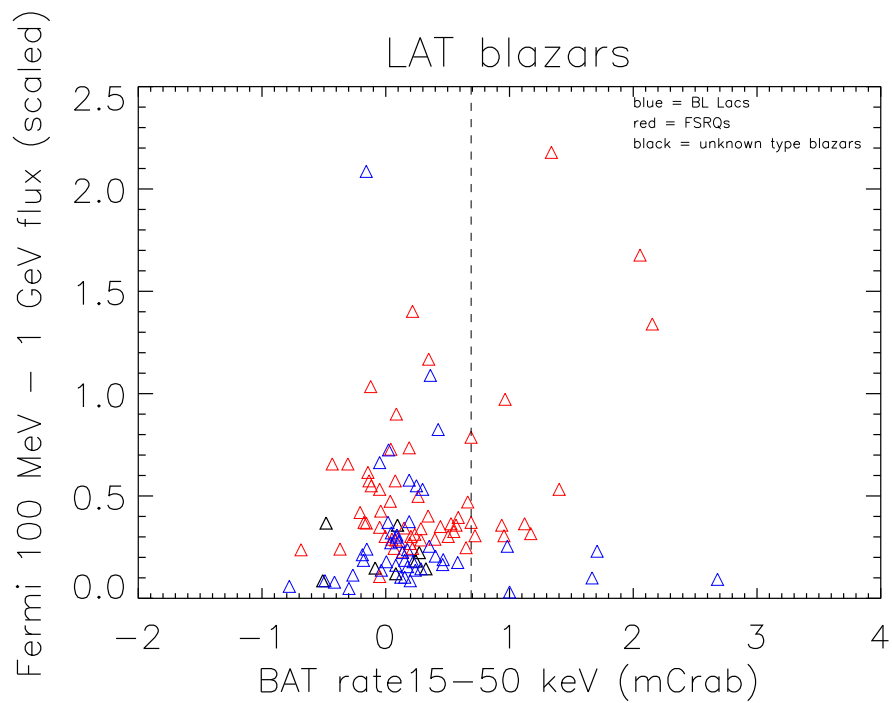} 
 \caption{The Fermi flux (scaled to the mean flux) in the 100 MeV -- 1 GeV band is plotted relative to the BAT Monitor count rate in mCrab.  The vertical line indicates our BAT detection threshold.  We see no apparent  correlation here, which is not surprising, given that the relationship between BAT and LAT flux depends strongly on where the peaks in the spectral energy distributions lie.
 }\label{fig3} 
\end{figure*}

\noindent Figure~\ref{fig3} shows that there is little or no correlation between the LAT 100 MeV to 1 GeV flux and the BAT 15-50 keV rate.  This is not surprising, since the relationship between these two quantities depends strongly on where the peak lies in the spectral energy distribution for each blazar.  A few points can be gleaned from this figure.  First of all, most of the BAT-detected BL Lacs (such as Mrk 421 and Mrk 501) are relatively faint in the LAT, while BAT tends to detect more of the brighter LAT flat-spectrum radio quasars.  Also, none of the unknown type blazars are detected in the BAT.

\section{Spectral Energy Distributions}\label{sed}

\begin{figure*}[t]
\centering
\includegraphics[width=135mm]{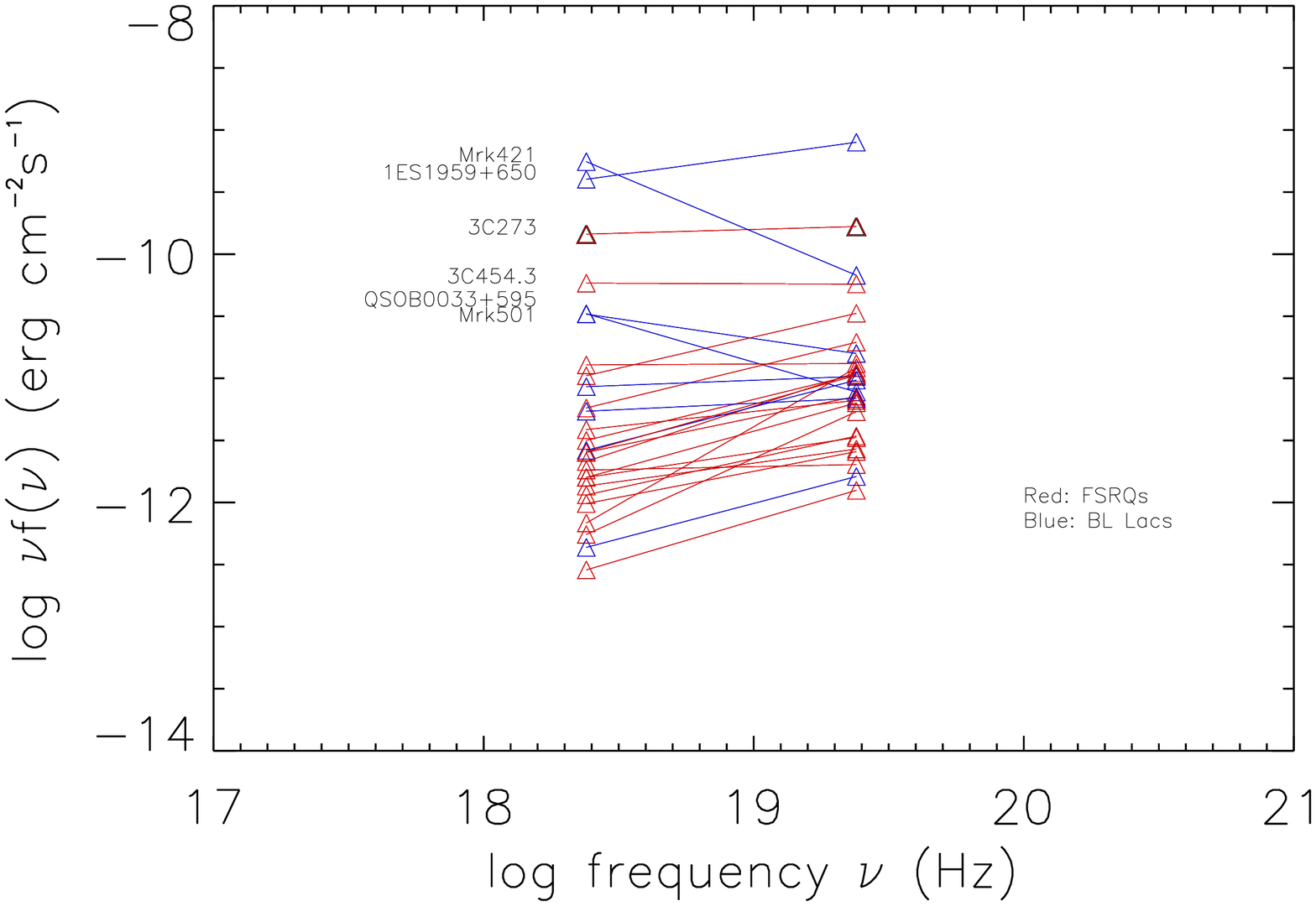} 
 \caption{Preliminary spectral energy distributions (SED) for BAT-detected blazars based on BAT and XRT observations.   }\label{fig4}
\end{figure*}

\noindent For 27 of the BAT-detected blazars, we have produced preliminary spectral energy distributions (SED) in the energy ranges of {\em Swift} XRT and BAT.   The XRT points were taken from a single {\em Swift} observation (the one with the longest exposure time) and the unabsorbed flux and photon index were derived using the web interface developed by Phil Evans at the UK Swift Science Data Centre at the University of Leicester\cite{evans09,pae-web}.   For four blazars (PKS 1830-211, PKS 0142-278, B3 2322+396 and PKS 2023-07) there were either no XRT observations in the archive or we were unable to adequately fit the spectrum.  The BAT SED points were derived using WebPIMMS (\verb+http://heasarc.gsfc.nasa.gov/Tools/w3pimms.html+) to convert from the BAT average 14-195 keV count rate to the flux density at 100 keV.  Thus Fig~\ref{fig4} does mix an XRT point at a single time with a mean BAT point.  Given the variations in blazar outputs, a more proper analysis (see Future Work) will include, where possible, more closely coincident spectral analysis.

For most flat-spectrum radio quasars (FSRQs), BAT appears to be on the rising part of the Compton peak, though for the brightest two (3C 273 and 3C 454.3), BAT is near the top of Compton peak.  For several of the high-frequency peaked BL Lacs (Mrk 421, Mrk 501, QSO B0033+595), the BAT appears to be on the falling edge of the synchrotron peak.  For others the dip in the SED is apparently below the BAT energy range, so the BAT points are again on the rising part of the Compton peak.

	\section {Future Work}

\begin{itemize}

\item {Re-run the BAT Transient Monitor to get complete light curves for all LAT bright sources back to the start of the {\em Swift} mission}
\item {Derive BAT spectral fits from the 8-channel BAT Survey results}
\item {Look for source time variability and either derive time-integrated XRT spectral fits or time-resolved BAT spectral fits}
\item {Look for temporal BAT-LAT correlations for the brightest BAT sources}
\item {Further study individual blazars and relate BAT results to what is known about particular sources and source classes.}

\end{itemize}

	\section{Data Tables}

\noindent The BAT results are shown in two tables.  In Table~\ref{tab1}, we list the 49 LAT sources which are detected in either the BAT monitor or BAT survey  with the criteria that {\em either} the source is seen at at least 0.64 mCrab in the BAT Transient Monitor or at $> 3\sigma$\ in the BAT Survey.  The common name and the source name in the {\em Fermi}/LAT Bright Source List are listed, along with the source type from the LAT\footnote{PSR=pulsar; rdg=radio galaxy; bzq=FSRQ blazar; bzb=BL Lac; x=special case; HXB=high-mass X-ray binary; NONE=no type listed in LAT source list}.  The results from the BAT Monitor are given in mCrab and from the BAT Survey in SNR.  The XRT flux and photon index were derived for the blazars (see Section~\ref{sed} for details).  

For two sources (noted with an asterisk), the LAT identification is with a region of the sky that contains the source listed: the Galactic Center region for Sgr A$\ast$ and the Large Magellanic Cloud for LMC X-1.  The BAT results are given for the individual sources, although there is still confusion at the BAT resolution for the Galactic Center.

One pulsar (PSR J2124-3358) and three of the four unidentified LAT sources in the table have BAT Survey SNR $< 0$, but a possible detection in the BAT Monitor.  This tells us that these sources are probably variable in the hard X rays, which is suggestive of their being galactic transient sources.

Table~\ref{tab2} shows the 20 brightest LAT sources overall and the 20 brightest blazars.  BAT-detected sources are indicated in bold and undetected sources in italics.  The LAT fluxes in the two bands are scaled to the average flux in each band and the rankings are based on the 100~MeV to 1~GeV flux. Only about half of the brightest LAT blazars are detected in the BAT, although most of the sources in the list have BAT Survey significance above $1.7\sigma$, so are likely to be detected in a deeper survey.

%


\begin{table}[t]
\begin{center}
\caption{Swift/BAT detections of Fermi/LAT bright sources}
\begin{tabular}{lllrrcc}
\hline  \textbf{Name} & \textbf{Fermi name} & \textbf{Type} & \textbf{ BAT monitor} &\textbf{BAT survey } & \textbf{ XRT flux} & \textbf{ XRT photon index}\\
 & & & \textbf{mCrab} & \textbf{SNR } & \textbf{ $erg\ cm^{-2} s^{-1}$} &  \\
\hline
Crab                      & J0534.6+2201 & PSR   & 1000.00 & 4605.02 & -- & --\\
Cen A                     & J1325.4-4303 & rdg   &   46.27 &  194.37 & -- & --\\
3C 273                    & J1229.1+0202 & bzq   &   11.28 &  101.78 & $2.90\times 10^{-10}$ & 1.64\\
$\ast$ Sgr Astar                 & J1746.0-2900 & NONE  &    7.79 &   32.87 & -- & --\\
Vela Pulsar               & J0835.4-4510 & PSR   &    5.73 &   45.46 & -- & --\\
Mrk 421                   & J1104.5+3811 & bzb   &    5.71 &   60.09 & $1.40\times 10^{-09}$ & 1.80\\
NGC 1275                  & J0320.0+4131 & rdg   &    4.51 &   30.88 & -- & --\\
3C 454.3                  & J2254.0+1609 & bzq   &    3.82 &   30.70 & $9.00\times 10^{-11}$ & 1.44\\
Mrk 501                   & J1653.9+3946 & bzb   &    2.68 &   16.23 & $1.40\times 10^{-10}$ & 2.11\\
$\ast$ LMC X-1                   & J0538.4-6856 & NONE  &    2.42 &   10.70 & -- & --\\
PKS 1830-211              & J1833.4-2106 & bzq   &    2.15 &    9.98 & -- & --\\
QSO J1512-0906            & J1512.7-0905 & bzq   &    2.05 &    9.08 & $1.20\times 10^{-11}$ & 1.17\\
QSO B0033+595             & J0036.7+5951 & bzb   &    1.71 &    8.67 & $1.30\times 10^{-10}$ & 2.06\\
1ES 1959+650              & J2000.2+6506 & bzb   &    1.67 &   13.41 & $3.50\times 10^{-10}$ & 2.46\\
PKS 2325+093              & J2327.3+0947 & bzq   &    1.40 &    6.05 & $3.50\times 10^{-12}$ & 1.49\\
QSO B1502+1041            & J1504.4+1030 & bzq   &    1.34 &    5.53 & $1.00\times 10^{-12}$ & 1.40\\
PSR J1418-60              & J1418.8-6058 & PSR   &    1.21 &    2.52 & -- & --\\
PKS 0142-278              & J0145.1-2728 & bzq   &    1.17 &    2.99 & -- & --\\
0FGL J1830.3+0617         & J1830.3+0617 & NONE  &    1.14 &   -0.56 & -- & --\\
RX J1826.2-1450           & J1826.3-1451 & hxb   &    1.14 &    3.86 & -- & --\\
QSO J1130-1449            & J1129.8-1443 & bzq   &    1.12 &   11.04 & $1.00\times 10^{-11}$ & 1.54\\
B3 2322+396               & J2325.3+3959 & bzb   &    1.00 &    0.87 & -- & --\\
QSO B2200+420             & J2202.4+4217 & bzb   &    0.98 &    8.09 & $2.90\times 10^{-11}$ & 1.98\\
0FGL J1834.4-0841         & J1834.4-0841 & x     &    0.96 &    1.52 & -- & --\\
3C 279                    & J1256.1-0547 & bzq   &    0.96 &    8.03 & $1.60\times 10^{-11}$ & 1.75\\
PMN J0948+0022            & J0948.3+0019 & bzq   &    0.96 &    1.38 & $3.80\times 10^{-12}$ & 1.67\\
PKS 1329-049              & J1331.7-0506 & bzq   &    0.94 &    4.19 & $3.40\times 10^{-12}$ & 1.69\\
PSR J1826-1256            & J1825.9-1256 & PSR   &    0.88 &    0.79 & -- & --\\
2CG 135+01                & J0240.3+6113 & HXB   &    0.77 &    9.88 & -- & --\\
0FGL J1231.5-1410         & J1231.5-1410 & NONE  &    0.76 &   -0.44 & -- & --\\
PSR J2124-3358            & J2124.7-3358 & PSR   &    0.75 &   -0.42 & -- & --\\
QSO J0217+0144            & J0217.8+0146 & bzq   &    0.72 &    2.11 & $3.30\times 10^{-12}$ & 1.67\\
0FGL J1413.1-6203         & J1413.1-6203 & NONE  &    0.71 &   -1.58 & -- & --\\
PKS 0528+134              & J0531.0+1331 & bzq   &    0.69 &    2.84 & $7.30\times 10^{-12}$ & 1.60\\
4C 11.69                  & J2232.4+1141 & bzq   &    0.69 &    6.66 & $6.60\times 10^{-12}$ & 1.67\\
PSR J1509-5850            & J1509.5-5848 & PSR   &    0.68 &    0.81 & -- & --\\
PSR J1809-2332            & J1809.5-2331 & PSR   &    0.67 &    0.88 & -- & --\\
PSR B1706-44              & J1709.7-4428 & PSR   &    0.67 &    2.45 & -- & --\\
PKS 0227-369              & J0229.5-3640 & bzq   &    0.66 &    1.75 & $1.60\times 10^{-12}$ & 1.49\\
PKS 1244-255              & J1246.6-2544 & bzq   &    0.65 &    1.65 & $2.50\times 10^{-12}$ & 2.46\\
QSO B2052-474             & J2056.1-4715 & bzq   &    0.57 &    3.45 & $4.90\times 10^{-12}$ & 1.62\\
QSO B0917+449             & J0921.2+4437 & bzq   &    0.50 &    3.05 & $5.10\times 10^{-12}$ & 2.04\\
QSO B1514-241             & J1517.9-2423 & bzb   &    0.46 &    4.51 & $5.70\times 10^{-12}$ & 1.70\\
QSO B0537-441             & J0538.8-4403 & bzb   &    0.36 &    5.73 & $1.40\times 10^{-11}$ & 1.82\\
QSO B2227-0848            & J2229.8-0829 & bzq   &    0.34 &    4.85 & $4.50\times 10^{-12}$ & 1.55\\
QSO B0716+714             & J0722.0+7120 & bzb   &    0.30 &    3.60 & $6.00\times 10^{-12}$ & 2.66\\
PKS 2023-07               & J2025.6-0736 & bzq   &    0.22 &    4.01 & -- & --\\
QSO B2201+1711            & J2203.2+1731 & bzq   &    0.19 &    3.13 & $1.10\times 10^{-12}$ & 1.64\\
8C 1849+670               & J1849.4+6706 & bzq   &    0.04 &    3.27 & $3.50\times 10^{-12}$ & 1.92\\
\hline
\end{tabular}
\label{tab1}
\end{center}
\end{table}

\begin{table}[t]
\begin{center}
\caption{The Brightest Fermi/LAT bright sources}
\begin{tabular}{lllrrc}
\hline \textbf{Name} & \textbf{Fermi name} & \textbf{Type} & \textbf{LAT Flux (scaled)} & \textbf{LAT flux (scaled)} & \textbf{BAT survey}\\
 & & & \textbf{(100 MeV -- 1 GeV} & \textbf{1 GeV -- 100 GeV} & \textbf{SNR}\\
{\normalsize Overall Brightest sources} & & & & & \\
\hline
{\bf Vela Pulsar              } & {\bf J0835.4-4510 } & {\bf PSR   } & {\bf 27.60} & {\bf 42.35} & {\bf   45.46} \\
{\it Geminga                  } & {\it J0634.0+1745 } & {\it PSR   } & {\it  9.84} & {\it 23.28} & {\it    0.97} \\
{\bf 3C 454.3                 } & {\bf J2254.0+1609 } & {\bf bzq   } & {\bf  7.28} & {\bf  3.71} & {\bf   30.70} \\
{\bf Crab                     } & {\bf J0534.6+2201 } & {\bf PSR   } & {\bf  7.01} & {\bf  5.82} & {\bf 4605.02} \\
{\it PSR J2021+4026           } & {\it J2021.5+4026 } & {\it PSR   } & {\it  4.25} & {\it  4.00} & {\it   -1.00} \\
{\bf Sgr Astar                } & {\bf J1746.0-2900 } & {\bf NONE  } & {\bf  4.03} & {\bf  2.99} & {\bf   32.87} \\
{\bf PSR B1706-44             } & {\bf J1709.7-4428 } & {\bf PSR   } & {\bf  3.96} & {\bf  5.98} & {\bf    2.45} \\
{\bf PSR J1826-1256           } & {\bf J1825.9-1256 } & {\bf PSR   } & {\bf  3.27} & {\bf  2.18} & {\bf    0.79} \\
{\it 0FGL J1855.9+0126        } & {\it J1855.9+0126 } & {\it x     } & {\it  3.06} & {\it  2.62} & {\it    0.80} \\
{\it 0FGL J1801.6-2327        } & {\it J1801.6-2327 } & {\it x     } & {\it  2.55} & {\it  1.70} & {\it   -0.22} \\
{\it PSR J2021+3651           } & {\it J2020.8+3649 } & {\it PSR   } & {\it  2.54} & {\it  2.37} & {\it    1.41} \\
{\bf 2CG 135+01               } & {\bf J0240.3+6113 } & {\bf HXB   } & {\bf  2.42} & {\bf  1.26} & {\bf    9.88} \\
{\bf RX J1826.2-1450          } & {\bf J1826.3-1451 } & {\bf hxb   } & {\bf  2.32} & {\bf  0.93} & {\bf    3.86} \\
{\bf 3C 273                   } & {\bf J1229.1+0202 } & {\bf bzq   } & {\bf  2.20} & {\bf  0.61} & {\bf  101.78} \\
{\bf QSO B1502+1041           } & {\bf J1504.4+1030 } & {\bf bzq   } & {\bf  2.18} & {\bf  2.21} & {\bf    5.53} \\
{\it AO 0235+16               } & {\it J0238.6+1636 } & {\it bzb   } & {\it  2.09} & {\it  2.57} & {\it    1.74} \\
{\it 0FGL J1813.5-1248        } & {\it J1813.5-1248 } & {\it NONE  } & {\it  1.85} & {\it  1.05} & {\it    0.08} \\
{\it 0FGL J1839.0-0549        } & {\it J1839.0-0549 } & {\it NONE  } & {\it  1.77} & {\it  1.53} & {\it   -0.62} \\
{\it PSR J2032+41             } & {\it J2032.2+4122 } & {\it PSR   } & {\it  1.76} & {\it  1.16} & {\it    1.88} \\
{\bf QSO J1512-0906           } & {\bf J1512.7-0905 } & {\bf bzq   } & {\bf  1.68} & {\bf  0.70} & {\bf    9.08} \\
\hline
{\normalsize Brightest blazars} & & & & & \\
\hline
{\bf 3C 454.3                 } & {\bf J2254.0+1609 } & {\bf bzq   } & {\bf  7.28} & {\bf  3.71} & {\bf  30.70} \\
{\bf 3C 273                   } & {\bf J1229.1+0202 } & {\bf bzq   } & {\bf  2.20} & {\bf  0.61} & {\bf 101.78} \\
{\bf QSO B1502+1041           } & {\bf J1504.4+1030 } & {\bf bzq   } & {\bf  2.18} & {\bf  2.21} & {\bf   5.53} \\
{\it AO 0235+16               } & {\it J0238.6+1636 } & {\it bzb   } & {\it  2.09} & {\it  2.57} & {\it   1.74} \\
{\bf QSO J1512-0906           } & {\bf J1512.7-0905 } & {\bf bzq   } & {\bf  1.68} & {\bf  0.70} & {\bf   9.08} \\
{\bf PKS 2023-07              } & {\bf J2025.6-0736 } & {\bf bzq   } & {\bf  1.40} & {\bf  0.83} & {\bf   4.01} \\
{\bf PKS 1830-211             } & {\bf J1833.4-2106 } & {\bf bzq   } & {\bf  1.34} & {\bf  0.39} & {\bf   9.98} \\
{\it PKS 0454-234             } & {\it J0457.1-2325 } & {\it bzq   } & {\it  1.17} & {\it  0.99} & {\it   1.77} \\
{\bf QSO B0537-441            } & {\bf J0538.8-4403 } & {\bf bzb   } & {\bf  1.09} & {\bf  0.95} & {\bf   5.73} \\
{\it PKS 1454-354             } & {\it J1457.6-3538 } & {\it bzq   } & {\it  1.03} & {\it  0.76} & {\it   2.11} \\
{\bf 3C 279                   } & {\bf J1256.1-0547 } & {\bf bzq   } & {\bf  0.97} & {\bf  0.54} & {\bf   8.03} \\
{\it QSO J0730-116            } & {\it J0730.4-1142 } & {\it bzq   } & {\it  0.90} & {\it  0.66} & {\it   1.74} \\
{\it 3C 66A                   } & {\it J0222.6+4302 } & {\it bzb   } & {\it  0.82} & {\it  0.99} & {\it   2.40} \\
{\bf PKS 0528+134             } & {\bf J0531.0+1331 } & {\bf bzq   } & {\bf  0.79} & {\bf  0.29} & {\bf   2.84} \\
{\it QSO B0208-5115           } & {\it J0210.8-5100 } & {\it bzq   } & {\it  0.74} & {\it  0.51} & {\it   1.15} \\
{\it TXS 1520+319             } & {\it J1522.2+3143 } & {\it bzq   } & {\it  0.73} & {\it  0.40} & {\it   2.15} \\
{\it PKS 0426-380             } & {\it J0428.7-3755 } & {\it bzb   } & {\it  0.73} & {\it  0.75} & {\it   2.14} \\
{\it PMN J1802-3940           } & {\it J1802.6-3939 } & {\it bzb   } & {\it  0.66} & {\it  0.56} & {\it  -0.62} \\
{\it B3 0650+453              } & {\it J0654.3+4513 } & {\it bzq   } & {\it  0.66} & {\it  0.48} & {\it  -0.46} \\
{\it PKS 1622-253             } & {\it J1625.8-2527 } & {\it bzq   } & {\it  0.66} & {\it  0.24} & {\it   1.59} \\

\end{tabular}
\label{tab2}
\end{center}
\end{table}


%





\end{document}